\documentclass[aip,pof,reprint,amsmath,amssymb]{revtex4-1}

\usepackage{amssymb,amsmath,amsfonts,graphicx}
\usepackage{bm}
\usepackage{comment}
\usepackage{subfigure}
\usepackage{xcolor}

\definecolor{red}{rgb}{1,0,0}
\definecolor{blue}{rgb}{0,0,1}

\newcommand{\beq}{\begin{equation}}
\newcommand{\eeq}{  \end{equation}}

\begin{document}

\preprint{AIP/123-QED}

\title{On the physics of fizzing: \\ How bubble bursting controls droplets ejection}

\author{Elisabeth Ghabache, Arnaud Antkowiak, Christophe Josserand, Thomas S\'eon}
\affiliation{Sorbonne Universit\'es, Universit\'e Pierre et Marie Curie and Centre National de la Recherche 
Scientifique, Unit\'e Mixte de Recherche 7190, Institut Jean Le Rond d'Alembert, 4 
Place Jussieu, F-75005 Paris, France}

\date{\today }

\begin{abstract}
\textbf{Abstract:} 
Bubbles at a free surface surface usually burst in ejecting myriads of droplets.
Focusing on the bubble bursting jet, prelude for these aerosols, we propose a simple scaling for the jet velocity and we unravel experimentally the intricate roles of bubble shape, capillary waves, gravity and liquid properties. We demonstrate that droplets ejection unexpectedly changes with liquid properties. 
In particular, using damping action of viscosity, self-similar collapse can be sheltered from capillary ripples and continue closer to the singular limit, therefore producing faster and smaller droplets. 
These results pave the road to the control of the bursting bubble aerosols.
\end{abstract}

\maketitle

%

\begin{figure*}[ht]
\centering
\noindent\includegraphics[width=\textwidth]{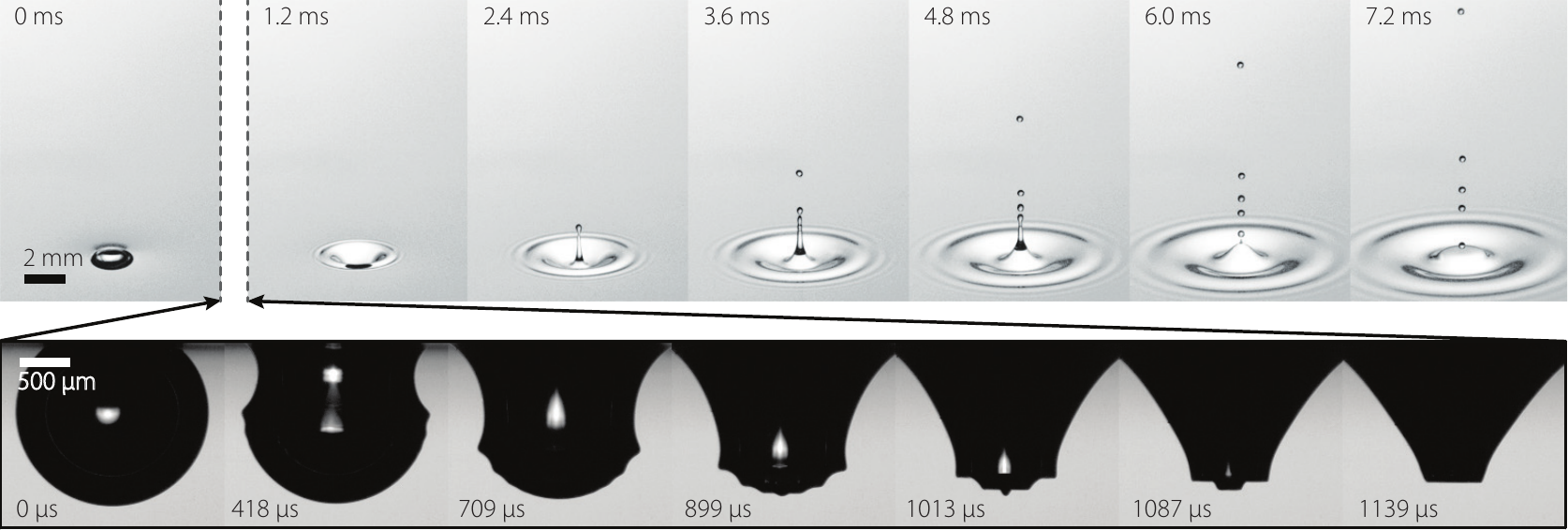}
\caption{Time sequence of a typical jetting event following a bubble bursting 
at a free surface in water. The top sequence shows the 
bubble bursting event above the free surface, while the bottom sequence displays the bubble under the free surface during
the collapse giving birth to the vertical upward jet. The bottom sequence takes place between the two first images of the top one. The times are shown on the snapshots with the same origin. See also the corresponding Videos S1 and S2.} 
\label{fig:sequence_jet}
\end{figure*}

Savoring a glass of champagne would not be as enjoyable without this fizzy sensation coming from bursting bubbles at the surface. 
More than just triggering a simple tingling sensation, the tiny droplets ejected during bursting are  crucial for champagne tasting as their evaporation highly contribute to the diffusion of wine aroma in air \cite{Liger-Belair2009}. 
Airborne droplets resulting from sea surface bubble bursting are also known since the late forties \cite{Woodcock1949, Woodcock1952, Woodcock1953} to play a major role in the interaction between ocean and atmosphere \cite{Andreas1995, Leeuw2011}. Two distinct types of droplets are involved, lying on two different mechanisms appearing during bubble bursting \cite{Blanchard1963}. 
When the thin liquid film - the bubble cap - separating the bubble from the atmosphere disintegrates, {\it film drops} are produced \cite{Knelman1954, lhuissier2012} with radius mainly less than 1\,$\mu$m. Then the resulting opened cavity (see Fig.\ref{fig:sequence_jet}) collapses and a jet emerges producing {\it jet drops} by breaking up \cite{Stuhlman1932, MacIntyre1972, Spiel1997}. For example, this latter mechanism accounts for the majority of sea-spray aerosol particles in the atmosphere with radius between 1 and 25 $\mu$m \cite{Lewis2004}.

The last sixty years have witnessed a number of laboratory studies documenting jet drops properties, such as the ejection speed, the maximum height or the size distribution as a function of bubble volume \cite{Hayami1958, Blanchard1963, Wu1973, Boulton-Stone1993, Spiel1995}, but a comprehensive picture of the mechanisms at play in bubble bursting is still lacking.
In particular, the sequence of violent events preluding jet formation \cite{MacIntyre1972, Duchemin2002} and the roles of liquid properties remain elusive. 

In this article, we unravel the tangled roles of liquid properties, gravity and capillary waves in the cavity collapse and show that these waves invariably adopt a self-similar behavior. 
We evidence the critical role of viscosity, that shelters self-similar collapse from remnant ripples, and therefore promotes the emergence of thinner and faster jets. Optimal conditions for singular jets as well as general scaling laws for the jet dynamics are assessed from detailed bubble bursting experiments. The consequences for aerosol generation are finally outlined, in particular in the context of champagne fizz, where liquid properties are tunable.


Our experiment consists in releasing a gas bubble from a submerged needle in a liquid and 
recording the upward jet after the bubble bursts at the free surface.
Bubbles are quasi-steadily formed using a syringe pump and detachment
frequency is weak enough to avoid successive bubbles interaction.
Different needle diameters (5 $<\Phi$~($\mu$m$)<$ 1800) allow 
us to create bubbles with various radii ($R$) ranging from 300 $\mu$m to 2000 $\mu$m.
The liquids used in this study 
include nine water-glycerol mixtures of viscosity in the range
$\mu$ = 1 mPa.s - 12 mPa.s, surface tension $\gamma$ = 64 mN.m$^{-1}$ - 72 mN.m$^{-1}$, 
and density $\rho$~=~1000~kg.m$^{-3}$ - 1160~kg.m$^{-3}$ and ethanol ($\mu$~=~1.2~mPa.s, $
\gamma$ = 23~mN.m$^{-1}$, $\rho$~=~780~kg.m$^{-3}$).
The height of fluid between needles and free surface is kept short (2-3 cm) to avoid 
rising bubble inflation. 
The bubble collapse and jet dynamics are analyzed through extreme close-up ultra-fast 
imagery. 
Macro lenses and extension rings allow us to 
record with a definition reaching 5 $\mu$m per pixel. Images are obtained 
between $10000$ and $150000$ frames per second using a digital high-speed camera 
(Photron SA-5). The ejection speed $V_\text{tip}$ is measured when the tip of the jet reaches the mean water level.

Figure~\ref{fig:sequence_jet}  (see also Videos S1 and S2)
illustrates a typical jetting event following a bubble bursting 
at a free surface in water. The top sequence shows the free surface view while the bottom one displays the underwater dynamics.
The first image of the top sequence shows a static bubble lying 
at the free surface. Then the film separating the bubble from the atmosphere drains and 
bursts leaving an unstable opened cavity. This cavity is millimeter-sized so the restoring 
force which tends to bring this hole back to a flat equilibrium is capillary driven. 
Bottom sequence displays capillary waves propagating along this cavity and focusing at the bottom. 
These collapsing waves give rise to a high speed vertical jet shooting out above the free surface as 
observed on the top sequence. The jet then fragments into droplets due to Rayleigh-Plateau 
destabilization generating an aerosol of one to ten droplets \cite{Blanchard1963}.


In order to establish the role played by the relevant 
parameters in the jet dynamics, we identify the five variables ruling the value of the jet tip velocity: 
$$V_\text{tip}=\Phi(R, \rho, \gamma, g, \mu)$$

\begin{figure}[ht]
\centering
\noindent\includegraphics[width=0.7\textwidth]{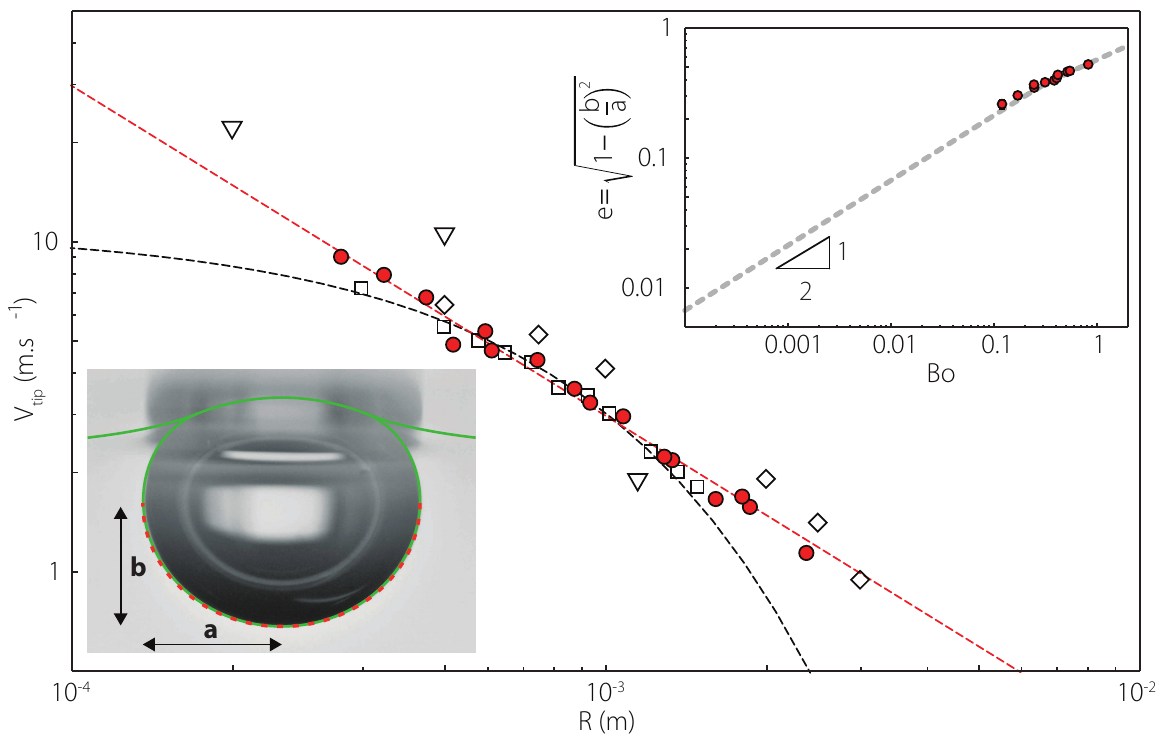}
\caption{Jet velocity $V_\text{tip}$ 
as a function of the bubble radius $R$ in water from our experiments (red circle) and data from the literature:  (square) top drop velocity from
Spiel \textit{et al.} \cite{Spiel1995} along with their exponential fit represented by the curved dashed line $V = 10.72\,\text{e}^{-1.27.10^3R}$ measured experimentally in fresh water,
(diamonds) maximum tip velocity computed numerically in fresh water from 
Boulton-Stone \textit{et al.}  \cite{Boulton-Stone1993} 
(triangle) top drop velocity from 
Blanchard \cite{Blanchard1963}. 
The red dashed line is a fit of our experimental data exhibiting the common trend $V_\text{tip} = \xi R^{-1}$ ($\xi = 2.95~10^{-3}$ m$^2$.s$^{-1}$). 
Bottom left inset: picture of a static floating bubble at the free surface. Green profile is obtained by numerical integration of the Young–Laplace equation using \textsc{Mathematica} software \cite{lhuissier2012, Toba1959}. The red dashed line is a semi-elliptic fit of the bottom part of the static bubble. $a$ and $b$ identify respectively the semi-minor and semi-major axis of the ellipsoid.
Top right inset: static bubble eccentricity $e = \sqrt{1-(a/b)^2}$ computed numerically as a function of the Bond number (dashed line). 
The red circles correspond to our experiments in water. Dashed line show asymptotic behaviors $e\propto \text{Bo}^{1/2}$ for small Bond number.} 
\label{fig:comp_modele}
\end{figure}

Using dimensional arguments, this equation becomes a relation between three 
dimensionless numbers fully describing the jet dynamics:
\begin{equation}
\label{ScalingLaw}
\text{We} = F(\text{Bo}, \text{Mo}),
\end{equation}
where  the Weber number $\text{We}= \rho V_\text{tip}^2 R/\gamma$ compares the effect of inertia and capillarity on the jet dynamics, the Bond number $\text{Bo} = \rho g R^2/\gamma$ compares the effect of gravity and capillarity on the initial bubble and  the Morton number $\text{Mo}~=~g \mu^4/\rho \gamma^3$ only depends on the fluid properties and is in particular independent of the bubble radius $R$. Various scaling relations for the velocity are reported in the literature, ranging from exponential fits of experimental data \cite{Spiel1995} (see curved dashed line Fig.~\ref{fig:comp_modele}) to algebraic laws $V_\text{tip} \propto R^{-1/2}$ in numerical simulations disregarding gravity \cite{Duchemin2002}. This diversity certainly emphasizes the need for further experimental analysis.

We set out by investigating experimentally in Fig.~\ref{fig:comp_modele} the dependence of $V_\text{tip}$ with $R$ in a log-log plot. Our experimental data (circles) rest along the line $V_\text{tip} \propto R^{-1}$, as indicated by the red dashed line fitting the experimental velocities. 
Note that bubbles with $\text{Bo} > 1$ (radii greater than 3mm) are out of scope of this study because they give rise to jets with a different dynamics and would constitute an other study.
On the same figure various data from the literature have been plotted: the top drop velocity measured experimentally in fresh water \cite{Spiel1995} or in sea water \cite{Blanchard1963}, and the maximum tip velocity of the jet computed numerically in fresh water \cite{Boulton-Stone1993}. 
It is noteworthy that our jet velocities match the first drop velocities (fitted by the exponential dashed line) making our results relevant for aerosol generation. 
Regardless of some slight differences they all follow the same trend $V_\text{tip} \propto R^{-1}$.
This specifies the form of the Eq.(\ref{ScalingLaw}) providing the variation with Bond number, yielding: 
\begin{equation}
\label{ScalingLaw2}
\text{We} = \text{Bo}^{-1/2} f(\text{Mo}).
\end{equation}
The $R^{-1}$ behavior is the footprint of gravity effects: the introduction of a second length scale, the gravity-capillary length $\ell_\text{gc} = \sqrt{\gamma/\rho g}$, allows departures from capillaro-inertial predictions through length scales ratios \cite{Seon2012}: $\text{We} = (\ell_\text{gc}/R) f(\text{Mo})$. Though the Bond number remains small in the experiments, the gravity plays a genuine role in the collapse dynamics that needs to be elucidated.
Froude number of the jet at the mean water level is $\text{Fr} = V_\text{tip}/\sqrt{gR}= \sqrt{\text{We}/\text{Bo}}$, and can be expressed here as $\text{Fr} = \text{Bo}^{-3/4} f(\text{Mo})^{1/2}$. In water, with the Bond values of Fig.\ref{fig:comp_modele} one obtains $\text{Fr} \in [7-160]>1$, indicating that gravity hardly affects the jet dynamics at least before eruption. 

Bottom left inset of Fig.~\ref{fig:comp_modele} displays a picture of the static bubble before bursting. Superimposed on the experimental picture, the profile is obtained by numerical integration of the Young–Laplace equation using \textsc{Mathematica} \cite{lhuissier2012, Toba1959}. The dashed line is a semi-elliptic fit of the bottom part of the static bubble allowing us to define the bubble parameters: $a$, $b$ and the corresponding bubble radius $R=(a^2b)^{1/3}$.
On the top right inset of Fig.\ref{fig:comp_modele} the eccentricity of the static floating bubble $e$ computed numerically is plotted versus the Bond number.
$e$ is found to decrease with the Bond number following $e\propto \text{Bo}^{1/2}$, demonstrating the non-sphericity of even small bubbles. This result naturally suggests that gravity influences the jet dynamics not through its direct action on the dynamics but by affecting the initial bubble shape.

\begin{figure}[ht]
\centering
\noindent\includegraphics[width=\textwidth]{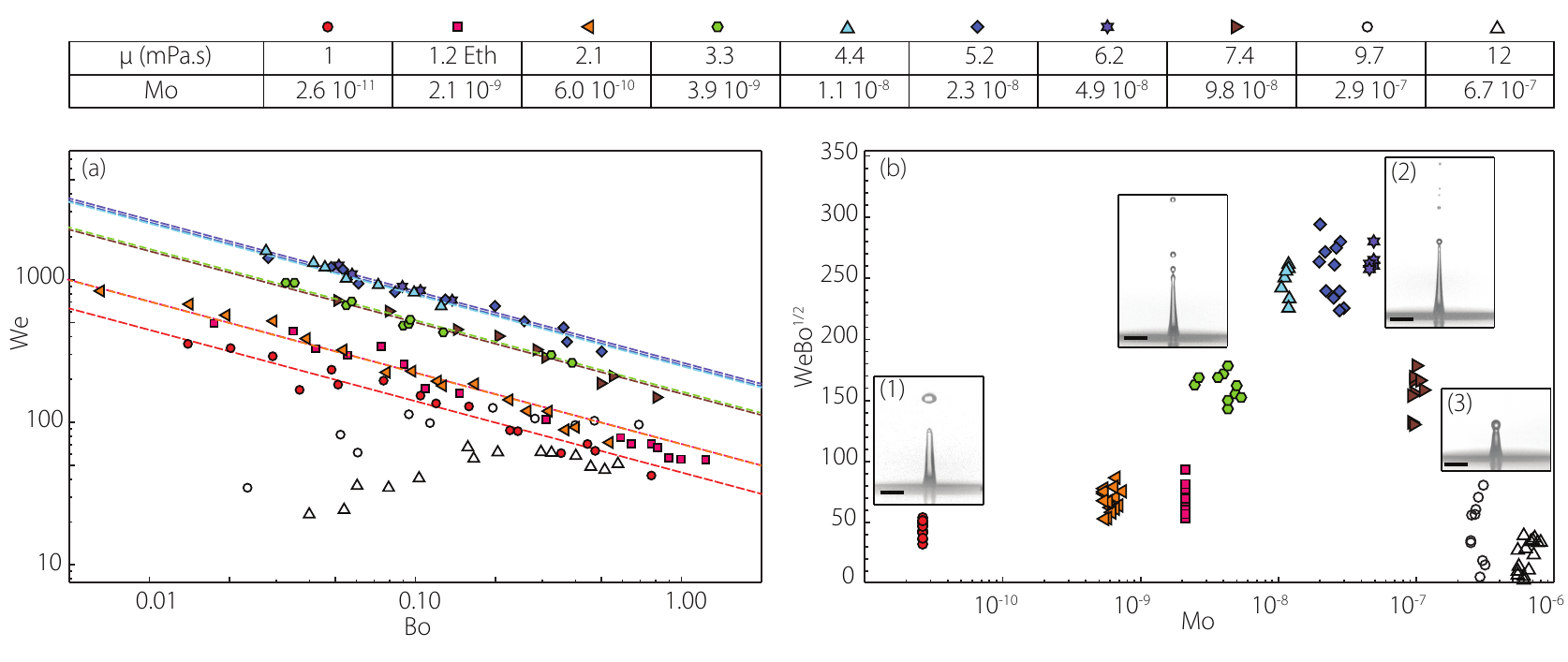}
\caption{Table: Value of the liquid viscosity and of the associated Morton number corresponding to each symbol. Eth. stands for Ethanol . 
(a) Weber number as a function of the Bond number for various values of the Morton number. All the colored symbols follow the same trend $\text{We} \propto \text{Bo}^{-1/2}$ as showed by the dashed lines. 
(b) $\text{We}\,\text{Bo}^{1/2}$ as a function of the Morton number $\text{Mo}$.  
Four snapshots display the typical jet observed at the same dimensionless time (t/$\sqrt{\rho R^3/\gamma}= 1/5 $), same Bond number ($\text{Bo}\simeq0.14$) and four different Morton numbers corresponding to red circle, green hexagon, purple star, empty triangle.  See also the Videos M1, M2 and M3 corresponding to the snapshots (1), (2) and (3). The black bar represents 500\,$\mu$m. } 
\label{fig:BellCurve}
\end{figure}

We now investigate how the jet eruption velocity $V_\text{tip}$ depends on the liquid properties and therefore on the Morton number. The Weber number is plotted as a function of the Bond number for various Morton number on Fig.~\ref{fig:BellCurve}(a). To browse the Morton range we mainly change the liquid viscosity. Correspondence between Morton number, liquid viscosity and symbols is indicated in the table of Fig.~\ref{fig:BellCurve}. The first clear observation is that the jet dynamics depends on the viscosity although the jet Reynolds number is greater than 1. 
Furthermore, the regime $\text{We} \propto \text{Bo}^{-1/2}$ is retained on around four decades in Morton number, from 1 mPa.s to around 7.5 mPa.s, all plotted with filled markers. 
This defines the boundary of our study considering that this viscous regime characterized by $\mu\gtrsim$ 9 mPa.s and showed with empty markers is out of the scope of this paper.  
Finally, for values of viscosity less than 6 mPa.s we observe a surprising increase of the Weber number with Morton number, meaning that for a given bubble radius in this range, {\it the jet is drastically faster  as the liquid viscosity is increased}. 

The non-dimensional jet velocity $\text{We}\, \text{Bo}^{1/2}$ is plotted as a function of the Morton number on Fig.~\ref{fig:BellCurve}(b), therefore specifying $f(\text{Mo})$ (see Eq.~\ref{ScalingLaw2}). A bell shaped curve is clearly observed with a maximum for $\mu = 5.2$ mPa.s.
To illustrate this unexpected behavior we display inside Fig.~\ref{fig:BellCurve} four snapshots of the jet at the same dimensionless time (t/$\sqrt{\rho R^3/\gamma}= 1/5 $), same Bond number ($\text{Bo}\simeq0.14$) but four different Morton numbers. The Videos M1, M2 and M3 correspond to the snapshots (1), (2) and (3).
The jet morphology undergoes a neat qualitative change as the liquid gets more viscous: the jet first becomes thinner, detaching more and smaller droplets and then ends up fat and small for high Morton number.

\begin{figure*}[ht]
\centering
\noindent\includegraphics[width=0.7\textwidth]{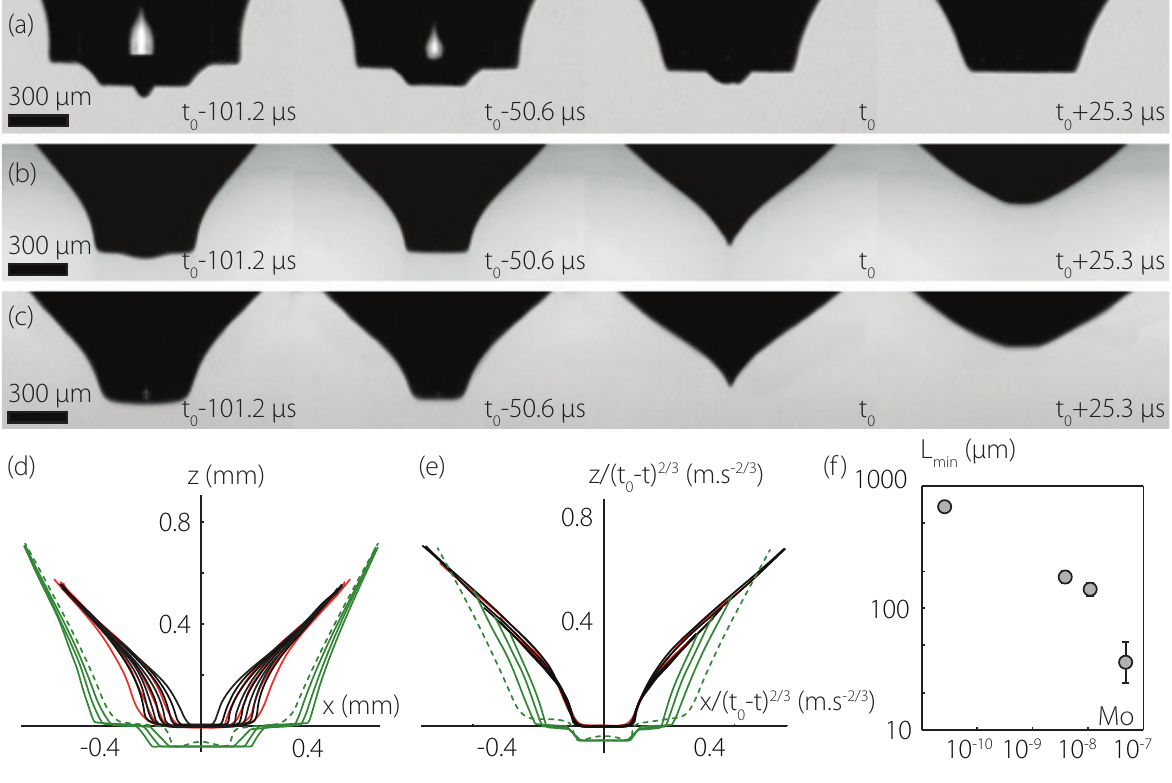}
\caption{Snapshots of the final stage of the cavity collapse before the jet develops. (a), (b) and (c) correspond respectively to the jets (1), (2) et (3) of Fig.~\ref{fig:BellCurve}. $t_0$ identifies the time of the wave collapse giving rise to the jet.
The cavity profiles corresponding to the three sequences are plotted at different times on (d) (green for sequence (a), black for (b) and red for (c)). 
(e) shows the collapse of these profiles according to the capillary-inertia self-similar behavior where lengths scale with $(t_0-t)^{2/3}$. (f)  L$_\text{min}$ versus Morton number. L$_\text{min}$ defines the width of the small left cavity when the collapse just quitted it self-similar behavior. The size of this effective cavity giving rise to the jet eruption decreases with viscosity for $\mu\lesssim$ 6 mPa.s.} 
\label{fig:Collapse}
\end{figure*}

In order to grasp the mechanisms leading to such a particular dynamics, we now turn to the jet formation by focusing on the cavity collapse {\it per se}. 
Lower sequence of Fig.~\ref{fig:sequence_jet} and Video S2 display a typical bubble collapse in water, where a train of capillary waves propagates, converges to the nadir (bottom of the cavity), and gives rise to the jet. 
Fig.\ref{fig:Collapse} shows a temporal zoom of the last microseconds before the cavity collapses ($t \simeq t_0$) for three different Morton numbers and same Bond number. These three sequences (a), (b) and (c) are the cavity collapse leading to the three jets (1), (2) and (3) displayed on Fig.~\ref{fig:BellCurve}. On the last image of each sequence the cavity is reversed and the upward jet (not seen on the picture) is developing. 
These sequences show that the cavity reversals are very similar between the $6.2$ and $12$ mPa.s solutions and drastically different from water.
In particular the small capillary waves present in the water collapse (a) have disappeared for higher viscosities (b) and  (c).
It has been shown in numerical simulation \cite{Duchemin2002} and in other experimental contexts \cite{zeff00,Bartolo06} that such collapse exhibits a self-similar dynamics that can lead in some cases to very thin and rapid jets. In such a situation, the cavity collapses through a nonlinear balance between capillary force and inertia, leading to a self-similar behavior where the lengths scale like $ (\gamma (t_0-t)^2/\rho)^{1/3~}$~\cite{Keller1983} ($t_0$ corresponding to the instant of the singular collapse) .
In the three cases presented here, the same self-similar collapse is clearly at play, as shown on 
Figs~\ref{fig:Collapse}(d,e) where the different cavity profiles plotted at different times before $t_0$ collapse when lengths are divided by $(t_0-t)^{2/3}$. 
So we observe a capillary-inertia self-similar collapse  for each case, even though the jets show clear differences (see Fig.~\ref{fig:BellCurve} (1), (2) and (3)) and are not singular.
Interestingly, the self-similar collapses for the high viscosity cases (b) and (c) are identical, consequently increasing the viscosity leads naturally to slow down the jet. However, these collapses are strongly different from the collapse in water (a) which is perturbed by the presence of the small capillary waves. These waves are always traveling on top of the interface and are inherent to the complex dynamics. But we observe that increasing the viscosity leads to smoothing the collapse.
In particular the closest time to $t_0$ in water, represented by the dashed profile, do not coalesce properly, signifying that the dynamics is no more self-similar.
This results in a collapse leaving its self-similar regime sooner than in a more viscous case, when the remnant ripples are damped.
By defining L$_\text{min}$ as the width of the small left cavity when the collapse just quitted it self-similar behavior, Fig.~\ref{fig:Collapse}(f) shows the variation of this effective collapsing cavity giving rise to the jet and reveals that it decreases with Morton number. This agrees with the idea of a self-similar collapse getting closer to the singularity as viscosity is increased and justifies why the jet velocity is increasing with Morton number.

This suggests an original mechanism to explain the role of the capillary waves for small Morton numbers. We consider that these waves break the self-similar dynamics when they are large enough (typically, when the wave amplitude is of the order of the self-similar structure).
Because the phase velocity of the capillary waves yields $ c \propto \sqrt{\gamma k/\rho}$, the shorter the wave, the faster it converges to the nadir, suggesting that the singular dynamics is destroyed by small waves first.
This picture has now to be corrected by the viscous damping of the capillary waves \cite{Zhang2008} which is also increasing with the wave number (with
damping rate $ \propto \mu k^2/\rho$). 
Therefore, as the viscosity increases, the interface is smoothed near the nadir and the instant where the oscillations destroy the self-similar dynamics is delayed closer and closer to the singularity.

Finally, two regimes of the jet dynamics as a function of the Morton number have been pointed out:  for $\text{Mo} \lesssim 3.10^{-8}$ the viscosity promotes the jet velocity by smoothing the collapsing cavity and  for $\text{Mo} \gtrsim 3.10^{-8}$ the jet velocity decreases with viscosity. At the frontier of these two regimes the jet is very thin and its velocity is maximal, which defines a region of the space phase where the aerosol production from bursting bubble is strongly enhanced.


The results presented in this paper apply for bubbles in newtonian fluids with liquid properties such that Bo $\in [10^{-2}-1]$ and  Mo $\in [10^{-11}-10^{-7}]$, which include most of the existing bursting bubble aerosols.
For instance, Bond and Morton numbers of champagne, from serving temperatures $4\,^{\circ}\mathrm{C}$ to room temperature, lie usually in the range $[8.10^{-3}-8.10^{-1}]$ $[6.10^{-10}-7.10^{-9}]$ respectively, where the droplets ejection sharply depends on the liquid properties (see Fig.\ref{fig:BellCurve}). 
These results are thus crucial in the context of champagne industry.
Indeed, quite recently \cite{Liger-Belair2009}, ultrahigh resolution mass spectrometry was used in order to analyze the aerosols released by bursting bubbles in champagne. In comparison with the bulk, champagne droplets were found to be over-concentrated with various surface active compounds, some of them showing indeed aromatic properties. This very characteristic fizz is therefore strongly believed to enhance the flavor sensation above a glass of bubbly wine in comparison with that above a glass of flat wine. 
It is now also well-known that specific treatment on champagne glass enables to create monodisperse bubbles reaching the surface at a chosen radius \cite{Liger-Belair2009a, Liger-Belair2013} and carboxymethyl cellulose (E466), used in food science as a  thickener, enables to modify the champagne viscosity with no consequences on the taste \cite{Bosso2010}. 
Therefore our results, by evidencing the existence of an unexpected maximum in the aerosols ejection speed and by providing this function between the first drop velocity and the bubble radius and liquid properties, pave the road to the characterization and control of the bursting bubble aerosols. They then constitute an important step forward to the fine tuning of champagne aroma diffusion, major goal of this industry. 
As an example, after determining the variation of the first droplet radius $r_d$ with Bond and Morton number and because we know its velocity, the vertical extension of the aerosols could easily be tuned. This will constitute a key result in the control of gaseous exchange between the aerosol and its surroundings. 
On the other hand, in the context of marine aerosols the Morton number ranges approximatively from 10$^{-11}$ to 10$^{-9}$, which is in the flat region of the jet velocity dependance on Mo, meaning that hydrodynamical properties of ocean, notably changing with temperature, barely affects the sea spray production. In this context quantity of results have been obtained characterizing the aerosols from bubble in water \cite{Lewis2004}. 
However, complementary studies on the number of droplets or drop size distribution as a function of bubble radius and liquid properties, based on the understanding of phenomena described here, will need to be realized to entirely characterize bursting bubble aerosols.

The Direction G\'en\'erale de l'Armement (DGA) is acknowledged for its financial support. We thank G\'erard Liger-Belair  for stimulating discussions about the relevance of our results in the champagne context, and for pointing out carboxymethyl cellulose as a possible viscosity modifier for champagne. 
We also thank Lucas Joseph for running the preliminary experiments.


\end{document}